# Structural modification in Au/Si(100) system: Role of surface oxide and vacuum level

A. Rath[1], J. K Dash[1], R. R. Juluri[1] and P. V. Satyam[1,*]

[1] Institute of Physics, Sachivalaya Marg, Bhubaneswar - 751005, India

**Abstract:**

To understand surface structural modifications for Au/Si (100) system, a thin gold film of ~2.0 nm was deposited under ultra high vacuum (UHV) condition on reconstructed Si surfaces using molecular beam epitaxy (MBE). Post annealing was done at 500°C in three different vacuum conditions: (1) low vacuum (LV) furnace ($10^{-2}$ mbar), (2) UHV ($10^{-10}$ mbar) (MBE chamber), (3) high vacuum (HV) chamber. The variation in the overall shape of the gold nanostructures and finer changes at the edges, like rounding of corners has been reported in this work. Although well aligned nano rectangles were formed in both HV and LV cases, corner rounding is more prominent in LV case. Furthermore in UHV case, random structures were formed having sharp corners. In all the above three cases, samples were exposed to air (for half an hour) before annealing. To study the effect of surface oxide, in-*situ* annealing inside UHV-MBE chamber was done without exposing to air. Well aligned rectangles with sharp corners (no corner rounding) were formed. The details about the role of surface oxides in the corner rounding process are discussed.

*Corresponding Author: satyam@iopb.res.in, pvsatyam22@gmail.com

# 1. Introduction

The spontaneous formation (self-assembly) of nano-objects during epitaxial growth offers the advantage of defect-free structures obtainable very quickly. However, knowledge of the fundamental processes governing crystal growth (microscopic pathways of diffusion, nucleation and aggregation) is essential in order to control the features of the nanostructures in a predictable way. On surfaces, the hierarchy in the migration barriers as well as the non uniform strain fields induced by mismatched lattice parameters can be translated into geometric order and well defined shapes and length scales of the resulting aggregates. Self-organizing nano crystal assemblies have already shown the degree of control necessary to address the challenges of building nanometer-scale technologies [1]. Metal nano-particles exhibit unique electronic, magnetic, photonic and catalytic properties which can be employed for the preparation of new materials for energy storage, photonics, communications, and sensing applications [2-7]. It is important to note that the average size distribution and interparticle separations of the Au NPs are very important for controlling the diameters and spacing in catalytic growth of nanowires [8].

However, controlling the shape of supported nano particles by thin film technologies is generally difficult since it depends both on kinetic and thermodynamic parameters [9]. However, this approach usually requires thermal activation through heating the support. Surface diffusion is a very important physical process that governs nano islands formation on surfaces. The control of the morphology (dimensions, shape, and number density) of nano islands is essential for exploiting their properties in many fields like optical applications or biosensors based on the use of surface plasmon resonances or nano catalysis in which the catalytic action depends on both shape of the nano islands and the nature of the substrate [10]. Well-defined shapes and uniform sizes are the most critical issues confronting the actual applications of these structures. Gold nano crystals with various shapes (rods, needles, squares etc) have been reported using various methods [11, 12]. Much progress has been made in recent years in the understanding of island shapes in epitaxial growth systems. Depending on fundamental growth parameters such as surface temperature and deposition rate, the deposited atoms form aggregates that may have a variety of shapes. Here we have reported the effect of vacuum conditions on morphology of the nano gold particles on Si(100) annealed at high temperature and how surface diffusion can be controlled in these processes.

In this paper, role of vacuum level has been studied on Au/Si(100) system without oxide at the interface. More precisely, effect of vacuum on surface diffusion and effect of surface oxides on the corner rounding have been studied. The aim of the work is to set an optimum growth and annealing condition to control the surface diffusion and hence the morphology of the gold structures for the real applications.

## 2. Experimental Details

A thin Au film of thicknesses 2.0 nm was deposited on n-type Si (100) of resistivity 10–20 Ω cm, by the molecular beam epitaxy (MBE) method under UHV (base pressure ≈ $2 \times 10^{-10}$ mbar) conditions [13]. The Si (100) substrates were loaded into the MBE chamber and degassed at ≈ 600°C for about 12 hours inside the chamber, followed by flashing for about 3 minutes by direct heating at a temperature of ≈1200°C. In this process, native oxide was removed and a clean reconstructed Si (100)-(2×1) surface (figure1 (a)) was obtained. On such ultra clean surfaces, gold films of different thicknesses were grown epitaxially by evaporating Au from a Knudshen cell. Deposition rate was kept constant at ≈ 0.14 nm min$^{-1}$. During the growth, the chamber vacuum was ≈ 6.2 ×10$^{-10}$ mbar. The thickness monitor was calibrated with RBS measurements. After the deposition, the samples were taken out of MBE chamber and annealed at 500$^0$C in three different vacuum conditions: (system-A) low vacuum (LV) external furnace (10$^{-2}$ mbar), (system-B) high vacuum (HV), (system-C) UHV (10$^{-10}$ mbar) chamber (MBE chamber). All above three cases, samples were exposed to air (for half an hour) before annealing. To study the effect of oxides we did the in-situ heating inside the UHV chamber without exposing to air (system-D). Planar TEM sample was prepared out of that annealed specimen for further TEM measurements with 200 keV electrons (2010, JEOL HRTEM).

## 3. Results and discussion

We have already shown that the as deposited MBE grown 2nm Au/Si(100) system (fig. 1(b) upon annealing inside the HV chamber (inside TEM) leads to formation of well aligned nano rectangular gold silicides structures[11] This shape transformation is attributed to the strain relief mechanism and detailed explanation is given in the reference [11]. The effect of vacuum level on the growth of above structures when there is no oxide at the interface has been discussed in this work. To study that, The MBE grown as-deposited samples were annealed at 500$^0$C in different vacuum conditions: HV, LV and UHV as mentioned in experimental details section. For HV (system- B)), well aligned nano rectangular structures were formed (fig. 2(a)). The length (longer side is always considered as length) distribution of the structures was shown in Fig. 2(b) and average length of the structures was 29.4 ± 1.5 nm. For LV (system- A), same well aligned nano rectangular structures were formed (fig. 2(c)) but corners of the rectangles were not sharp. Due to the different bonding strengths along two facets the migration barrier experienced by the gold atoms along the length of the rectangle and around the corners differ, which causes the rounding of the corners [14]. The corner rounding effect is more prominent compare to the system-B. Average length of the structures was 21.9 ± 1.2 nm (figure 2(d)). There was no much change in mean length between system-A and system-B. Whereas for UHV case (system-C), there was increase in mean length (fig. 3(a)) and the structures were not full rectangular like previous case. The average length of the structure was 38.2 ± 1.4 nm (fig. 3(b)). But it is interesting to notice that the corners are very sharp (fig. 3(a)). It should be noted that in all above cases, samples were exposed to air (for half an hour) before doing the annealing experiment.

To study the effect of oxides we did the in-situ heating at 500°C inside UHV chamber without exposing to air (system-D). Well aligned rectangles with sharp corners (no corner rounding) were formed (fig. 3(c)). In fig. 3(d) the length distribution shows broad distribution having average length 56.4 ± 1.6 nm. Upon air exposure, the native oxide was grown on the inter island regions, which hinder the surface reconstruction; but the Au-Si islands were still intact to the reconstructed 2×1 surface. During UHV annealing (system-C), the islands follow the substrate symmetry due to the absence of native oxide layer at the Au-Si interface. Whereas oxide desorption starts in the inter island regions, creating unsaturated bonds, which have a great affinity to attract the nearest gold particles. Therefore, the Au-Si islands diffuse towards the oxide desorbed region to minimize the surface energy, which results in the formation of irregular structures. But in contrast to the above case, in the *in-situ* UHV annealing (without surface oxide layer, (system-D)), the Au-Si islands form a regular rectangular shape following the 2×1 reconstruction.

As mentioned earlier, in LV (system-A) and HV (system-B) annealing case, well aligned rectangular structures were formed on the same air exposed samples unlike in the UHV annealing case (system-C). It has been reported that at LV, HV and UHV, the redeposition of oxide takes about 2.0 ms, 2.0 s and 1 hour respectively to form a monolayer [15-17]. Thus, if this rate of redeposition of oxide exceeds the rate of decomposition, proper desorption might not be possible. As the oxide desorption is more in UHV compare to HV and LV, the diffusion of Au-Si islands towards the desorbed area is less probable in later cases. So the islands get less deformed in HV and LV annealing, forming a regular stable shape.

## 4. Conclusions

2.0 nm Au samples were deposited on a cleaned Si (100) by MBE method and thermally annealed in different vacuum level (LV, HV and UHV). Aligned nanostructures were observed in all three cases. The variation in the overall shape of the gold nanostructures and finer changes at the edges, like rounding of corners has been demonstrated in this work. This can be attributed to the variation in diffusion of gold silicide islands due to the presence of surface oxides. These near-regularly spaced gold silicide nano rectangles on a silicon substrate could serve as a simple chemically reactive template for potential applications in nano electronics and catalysts.

**Figure Captions**

**Fig 1**(a) Reflection high energy electron diffraction (RHEED) pattern showing clean reconstructed Si100-(2×1) surface (b) Bright field TEM image shows the as deposited 2nm Au/Si(100) system.

**Fig 2** Bright field TEM image of MBE grown 2nm Au/Si(100) system annealed upto 500°C inside (a) HV chamber and (c) LV chamber having average length 29.4 ± 1.5 nm (b) and 21.9 ± 1.2 nm (c) respectively. Inset figure shows the HRTEM image of one of the structures

**Fig 3** Bright field TEM image of MBE grown 2nm Au/Si(100) system annealed upto 500°C inside UHV chamber (a) (with air-exposed) and (c) without air exposed (in-situ) having average length 38.2 ± 1.4 nm (b), 56.4 ± 1.6 nm (d) respectively. Inset figure shows the HRTEM image of one of the structures

Fig 1: Rath et al

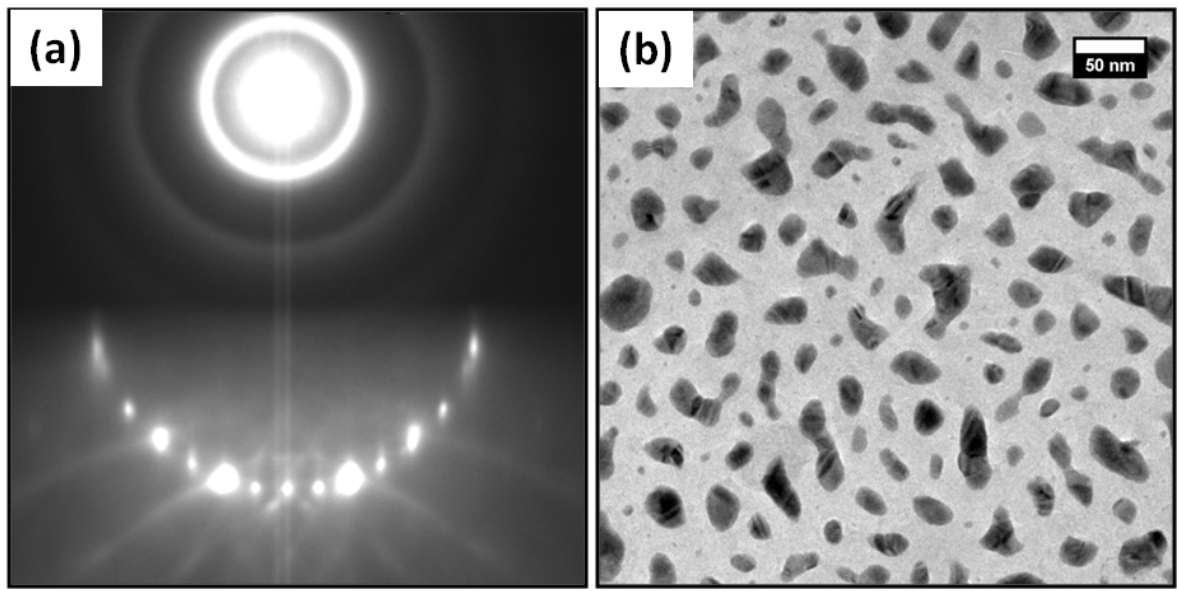

**Fig 2: Rath et al**

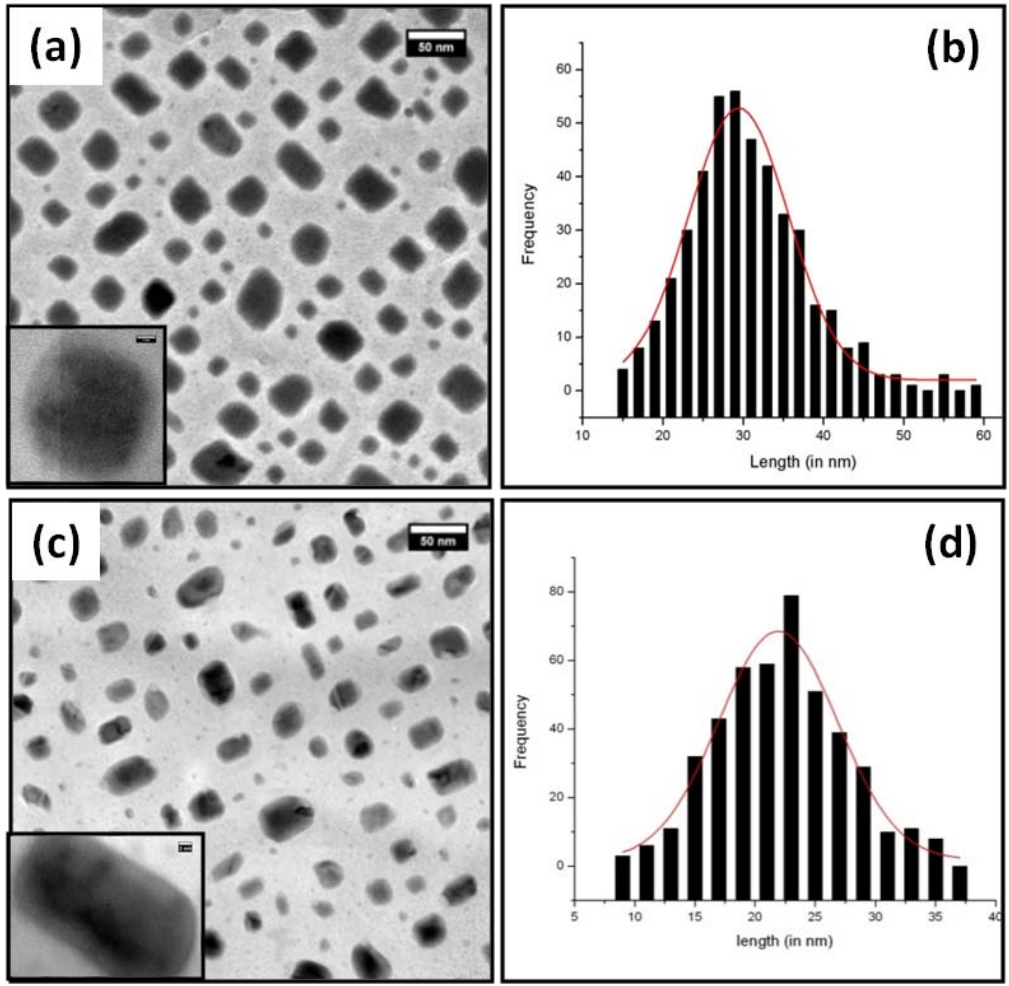

**Fig 3; Rath et al**

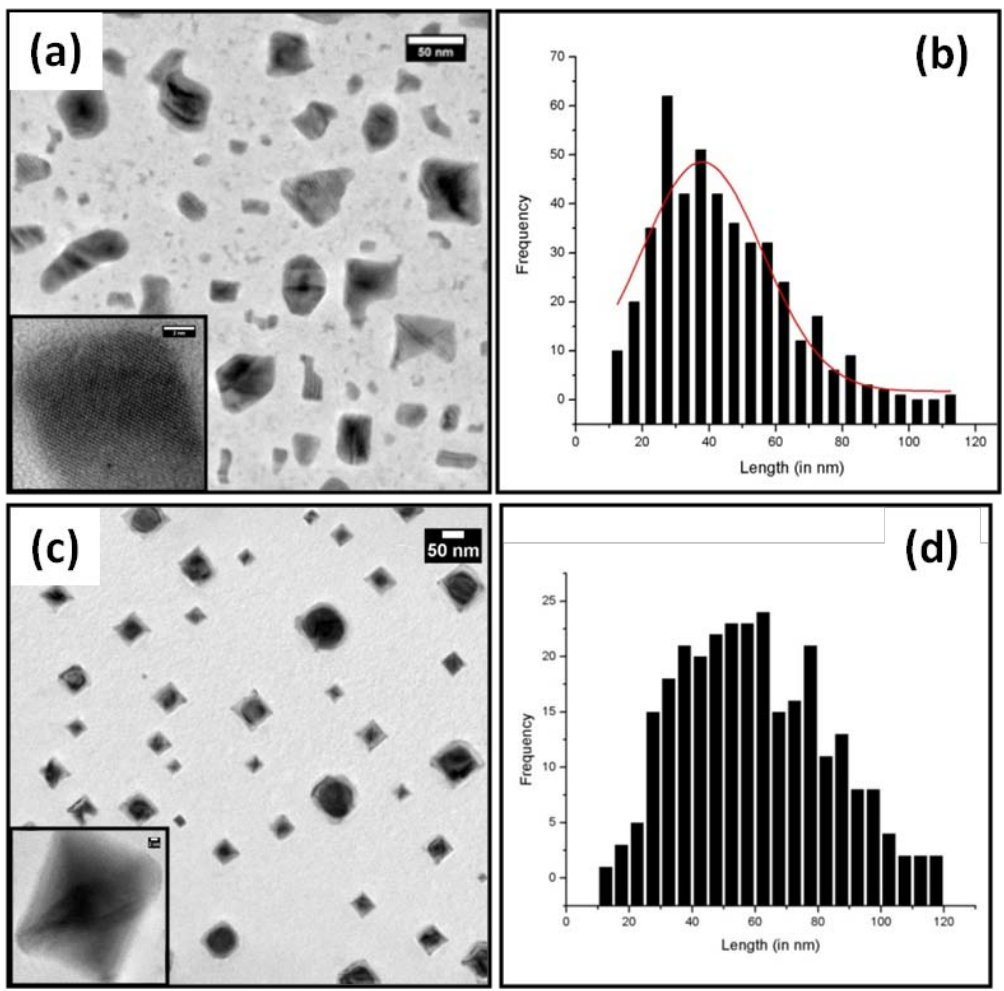